\documentclass[a4paper,10pt]{scrartcl}
\usepackage[T1]{fontenc}
\usepackage[english]{babel}
\usepackage{amsmath}
\usepackage{amssymb}
\usepackage{amsfonts}
\usepackage{stmaryrd}
\usepackage[utf8]{inputenc}
\usepackage{natbib}

\begin{document}

\author{T. Hartung}
\title{On radial migration of dense regions and objects and local unstability of accreting systems}
\maketitle

\begin{abstract}
I have used a newtonian infinite body problem to model a protoplanitary-like hot accretion disk and added terms of laminar and Stokes friction. Then I used qualitative methods to show the unstability of local regions leading to a lemma of radial migration and local unstability in hot, proto-planetary-like accretion disks. Then I considered a general infinite body problem with a position dependent, gravitation-like force and a general perturbating term.
\end{abstract}

\section*{Introduction}
\paragraph{}
In (\citet{har}) I derived an $\infty$-body model of a protoplanitary-like disk mainly considering the barycenter of the disk itself and a spherical local region, the observed object is located in, with its radius and barycenter. The Lagrangian of a dense region in this model is
\begin{equation}
 \mathcal{L}=\frac{2(T-U)}{m_3}=\dot x^2+\dot y^2+\frac{\tilde\alpha_1}{\left(x^2+y^2\right)^{\frac{1}{2}}}+\tilde\alpha_2\left(\frac{(x-\chi)^2+(y-\psi)^2}{R^2}-3\right)
\end{equation}
with $m_3$ being the mass of the object, $x$ and $y$ the coordinates with respect to the barycenter of the disk, $\chi$ and $\psi$ the coordinates of the local barycenter with respect to the barycenter of the disk, $R$ the radius of the local region and $\alpha_i$ giving the strength of the accreting force. This model has shown to give an explanation to the observations described by \citet{ros}.

\section*{Changes to the model}
\paragraph{}
For now it will be interesting to see how the results will change using friction. Therefore I will consider laminar and Stokes friction. As for laminar friction the equation of motion reads
\begin{equation}
 \left(\begin{array}{c}
\ddot x\\ \ddot y
\end{array}
\right)=\left(\begin{array}{c}
\frac{\alpha_1x}{\left(x^2+y^2\right)^{\frac{3}{2}}}+\alpha_2\left(x-\chi\right)-\kappa_l\dot x\\
\frac{\alpha_1y}{\left(x^2+y^2\right)^{\frac{3}{2}}}+\alpha_2\left(y-\psi\right)-\kappa_l\dot y
\end{array}
\right)
\end{equation}
and for Stokes friction it reads
\begin{equation}
 \left(\begin{array}{c}
\ddot x\\ \ddot y
\end{array}
\right)=\left(\begin{array}{c}
\frac{\alpha_1x}{\left(x^2+y^2\right)^{\frac{3}{2}}}+\alpha_2\left(x-\chi\right)-\kappa_s\dot x^2\\
\frac{\alpha_1y}{\left(x^2+y^2\right)^{\frac{3}{2}}}+\alpha_2\left(y-\psi\right)-\kappa_s\dot y^2
\end{array}
\right)
\end{equation}
with $\kappa_{l,s}>0$. Thus the systems are
\begin{equation}
\left(\begin{array}{l}
 \dot x\\
\dot v_x\\
\dot y\\
\dot v_y
\end{array}
\right)=  f_l\left(
x,
v_x,
y,
v_y
\right):=\left(\begin{array}{c}
 v_x\\
\frac{\alpha_1x}{\left(x^2+y^2\right)^{\frac{3}{2}}}+\alpha_2\left(x-\chi\right)-\kappa_l v_x\\
v_y\\
\frac{\alpha_1y}{\left(x^2+y^2\right)^{\frac{3}{2}}}+\alpha_2\left(y-\psi\right)-\kappa_l v_y
\end{array}
\right)
\end{equation}
and
\begin{equation}
\left(\begin{array}{l}
 \dot x\\
\dot v_x\\
\dot y\\
\dot v_y
\end{array}
\right)=  f_s\left(
x,
v_x,
y,
v_y
\right):=\left(\begin{array}{c}
 v_x\\
\frac{\alpha_1x}{\left(x^2+y^2\right)^{\frac{3}{2}}}+\alpha_2\left(x-\chi\right)-\kappa_s v_x^2\\
v_y\\
\frac{\alpha_1y}{\left(x^2+y^2\right)^{\frac{3}{2}}}+\alpha_2\left(y-\psi\right)-\kappa_s v_y^2
\end{array}
\right)
\end{equation}

\section*{Stability of local trajectories}
\paragraph{}
To retrieve informations on the stability I will search for Lyapunovfunctions $V_{l,s}$ of both systems. They will have to integrate the pde's
\begin{multline}\label{dVl}
 0\ge\dot V_l:=\langle \nabla V_l|f_l\rangle=v_x\partial_xV_l+\left(\frac{\alpha_1x}{\left(x^2+y^2\right)^{\frac{3}{2}}}+\alpha_2\left(x-\chi\right)-\kappa_l v_x\right)\partial_{v_x}V_l\\
+v_y\partial_yV_l+\left(\frac{\alpha_1y}{\left(x^2+y^2\right)^{\frac{3}{2}}}+\alpha_2\left(y-\psi\right)-\kappa_l v_y\right)\partial_{v_y}V_l
\end{multline}
and
\begin{multline}\label{dVs}
 0\ge\dot V_s:=\langle \nabla V_s|f_s\rangle=v_x\partial_xV_s+\left(\frac{\alpha_1x}{\left(x^2+y^2\right)^{\frac{3}{2}}}+\alpha_2\left(x-\chi\right)-\kappa_s v_x^2\right)\partial_{v_x}V_s\\
+v_y\partial_yV_s+\left(\frac{\alpha_1y}{\left(x^2+y^2\right)^{\frac{3}{2}}}+\alpha_2\left(y-\psi\right)-\kappa_s v_y^2\right)\partial_{v_y}V_s
\end{multline}
For Stokes it's easy to see, that
\begin{equation}\label{Vs}
V_s:=\begin{cases}\begin{array}{ll}
\exp\left(-x'^2+v_x'^2-y'^2+v_y'^2\right)-1&,\ x'\ge0,\ v_x'\ge0,\ y'\ge0,\ v_y'\ge0\\
\exp\left(-x'^2+v_x'^2-y'^2-v_y'^2\right)-1&,\ x'\ge0,\ v_x'\ge0,\ y'\ge0,\ v_y'\le0\\
\exp\left(-x'^2+v_x'^2+y'^2+v_y'^2\right)-1&,\ x'\ge0,\ v_x'\ge0,\ y'\le0,\ v_y'\ge0\\
\exp\left(-x'^2+v_x'^2+y'^2-v_y'^2\right)-1&,\ x'\ge0,\ v_x'\ge0,\ y'\le0,\ v_y'\le0\\
\exp\left(-x'^2-v_x'^2-y'^2+v_y'^2\right)-1&,\ x'\ge0,\ v_x'\le0,\ y'\ge0,\ v_y'\ge0\\
\exp\left(-x'^2-v_x'^2-y'^2-v_y'^2\right)-1&,\ x'\ge0,\ v_x'\le0,\ y'\ge0,\ v_y'\le0\\
\exp\left(-x'^2-v_x'^2+y'^2+v_y'^2\right)-1&,\ x'\ge0,\ v_x'\le0,\ y'\le0,\ v_y'\ge0\\
\exp\left(-x'^2-v_x'^2+y'^2-v_y'^2\right)-1&,\ x'\ge0,\ v_x'\le0,\ y'\le0,\ v_y'\le0\\
\exp\left(x'^2+v_x'^2-y'^2+v_y'^2\right)-1&,\ x'\le0,\ v_x'\ge0,\ y'\ge0,\ v_y'\ge0\\
\exp\left(x'^2+v_x'^2-y'^2-v_y'^2\right)-1&,\ x'\le0,\ v_x'\ge0,\ y'\ge0,\ v_y'\le0\\
\exp\left(x'^2+v_x'^2+y'^2+v_y'^2\right)-1&,\ x'\le0,\ v_x'\ge0,\ y'\le0,\ v_y'\ge0\\
\exp\left(x'^2+v_x'^2+y'^2-v_y'^2\right)-1&,\ x'\le0,\ v_x'\ge0,\ y'\le0,\ v_y'\le0\\
\exp\left(x'^2-v_x'^2-y'^2+v_y'^2\right)-1&,\ x'\le0,\ v_x'\le0,\ y'\ge0,\ v_y'\ge0\\
\exp\left(x'^2-v_x'^2-y'^2-v_y'^2\right)-1&,\ x'\le0,\ v_x'\le0,\ y'\ge0,\ v_y'\le0\\
\exp\left(x'^2-v_x'^2+y'^2+v_y'^2\right)-1&,\ x'\le0,\ v_x'\le0,\ y'\le0,\ v_y'\ge0\\
\exp\left(x'^2-v_x'^2+y'^2-v_y'^2\right)-1&,\ x'\le0,\ v_x'\le0,\ y'\le0,\ v_y'\le0\\
\end{array}\end{cases}
\end{equation}
with $x-\chi=:x'$, $v_x-v_\chi=:v_x'$, $y-\psi=:y'$ and $v_y-v_\psi=:v_y'$ also integrates (\ref{dVs}) on 
\begin{equation}
 \Omega_s:=\left\{(x,v_x,y,v_y)^T;\ |x-\chi|<\frac{\alpha_1x}{\alpha_2(x^2+y^2)^{\frac{3}{2}}},\ |y-\psi|<\frac{\alpha_1y}{\alpha_2(x^2+y^2)^{\frac{3}{2}}}\right\}
\end{equation}
with $V_s(\chi,v_\chi,\psi,v_\psi)=0=\dot V_s(\chi,v_\chi,\psi,v_\psi)$ and coordinates respectively chosen to comply $x,y,v_x,v_y,\chi,\psi>0$, since $\kappa_s v_x^2>0<\kappa_s v_y^2$ and thus $\dot V_s\le \dot V\le0$ (using $V$ from (\citet{har})). Reducing $\Omega_s$ to
\begin{multline}
 \Omega_l:=\left\{(x,v_x,y,v_y)^T;\ |x'|<\left\lvert\kappa_lv_x-\frac{\alpha_1x}{\alpha_2(x^2+y^2)^{\frac{3}{2}}}\right\rvert,\right.\\ \left.|y'|<\left\lvert\kappa_lv_y-\frac{\alpha_1y}{\alpha_2(x^2+y^2)^{\frac{3}{2}}}\right\rvert\right\}
\end{multline}
$V_l=V_s$ still holds (\ref{dVl}). Obviously $\Omega_l$ as well as $\Omega_s$ are depending on the coordinates chosen. In case the chosen system gives $\Omega_{l,s}=0$ simply rotate the system. This will always be possible (see \citet{har}). In any given system one will be able to estimate the global constrained supremum of $R$ holding 
\begin{equation}
B((\chi,\psi,v_\chi,v_\psi,\chi,\psi,v_\chi,v_\psi);R)\cap\Omega_{l,s}=B((\chi,\psi,v_\chi,v_\psi,\chi,\psi,v_\chi,v_\psi);R)
\end{equation}
$B(a;r)$ be the open ball with center $a$ and radius $r$. $B((\chi,\psi,v_\chi,v_\psi,\chi,\psi,v_\chi,v_\psi);R)$ then is the largest possible local region to be considered.

\section*{Summerization}
\paragraph{}
Recapitulating the calculations above and (\citet{har}) they can be summerized as a
\paragraph{} \textbf{Lemma 1}\\
Let $X:=(x,y,v_x,v_y,\chi,\psi,v_\chi,v_\psi)\in\mathbb{X}$, with 
\begin{multline*}
\mathbb{X}:=\{X;\ x,y,v_x,v_y,\chi,\psi>0,\\ \operatorname{arg}(x+iy)-\operatorname{arg}(x_0+iy_0)=\operatorname{arg}(\chi+i\psi)-\operatorname{arg}(\chi_0+i\psi_0)=\varphi\in\mathbb{R}\}
\end{multline*}
and $X_0:=(x_0,y_0,v_{x_0},v_{y_0},\chi_0,\psi_0,v_{\chi_0},v_{\psi_0})$ the coordinates in any given system of coordinates, be the center-of-mass coordinates of a dense object/region $D$ ($x,y,v_x,v_y$) and its local surrounding region $L$ ($\chi,\psi,v_\chi,v_\psi$) with respect to the barycenter of a hot, protoplanetary-like accretion disk $A$. Let $D$ be in the middle region of $A$. Furthermore let $\alpha_1$ be the effective gravitational constant of the center-of-mass of $A$, let $\alpha_2$ be the effective gravitational constant of the center-of-mass of $L$ and let $\kappa$ be the effective dissipative, velocity proportional force parameter, then $D$ will radially migrate through $A$ and $A$ is unstable on resolution lengthscales $R\ge r$ holding
\begin{displaymath}
 r:=\sup\{\varepsilon>0;\ \exists \varphi\in\mathbb{R}:\ B((L,L);\varepsilon)\cap\Omega_\varphi=B((L,L);\varepsilon)\}
\end{displaymath}
with
\begin{displaymath}
 \Omega_\varphi:=\left\{X\in\mathbb{X}\arrowvert_{\varphi};\ |x-\chi|<\left\lvert\kappa v_x-\frac{\alpha_1x}{\alpha_2(x^2+y^2)^{\frac{3}{2}}}\right\rvert,\ |y-\psi|<\left\lvert\kappa v_y-\frac{\alpha_1y}{\alpha_2(x^2+y^2)^{\frac{3}{2}}}\right\rvert\right\}
\end{displaymath}

\section*{Generalization}
\paragraph{}
For consideration of more general scenarios, I will still use the $\infty$-body problem with the same definition of phasespace-coordinates. But the system shall be altered to three dimensions in space and considering a general gravitation-like force $F=(F_x,F_y,F_z)^T(x,y,z)$ and a perturbating term \\$G=(G_x,G_y,G_z)^T(x,y,z,v_x,v_y,v_z,\chi,\psi,\zeta,v_\chi,v_\psi,v_\zeta)$. Thus the system reads
\begin{equation}
 \left(\begin{array}{c}
 \dot x\\\dot y\\\dot z\\\dot v_x\\\dot v_y\\\dot v_z
\end{array}\right)= \left(\begin{array}{c}
 v_x\\v_y\\v_z\\F_x+G_x\\F_y+G_y\\F_z+G_z
\end{array}\right)
\end{equation}
A Lyapunovfunction $V$ needs to hold
\begin{multline}\label{dV}
 0\ge \dot V:=v_x\partial_xV+v_y\partial_yV+v_z\partial_zV\\+(F_x+G_x)\partial_{v_x}V+(F_y+G_y)\partial_{v_y}V+(F_z+G_z)\partial_{v_z}V
\end{multline}
Defining primed coordinates analogous to $x':=x-\chi$ it is possible to define $V$ like (\ref{Vs})
\begin{multline}
V:=\exp\left(-\operatorname{sgn}(x'v_x)x'^2-\operatorname{sgn}(y'v_y)y'^2-\operatorname{sgn}(z'v_z)z'^2-\operatorname{sgn}(v_x'(F_x+G_x))v_x'^2\right.\\
-\left.\operatorname{sgn}(v_y'(F_y+G_y))v_y'^2-\operatorname{sgn}(v_z'(F_z+G_z))v_z'^2\right)-1
\end{multline}
Since coordinate functions can be chosen to just be zero at discrete times and finitely often on open timeintervals, they do not have any impact on the stability. Furthermore accreting systems need to hold $(F+G)<0$ at most times. $(F+G)>0$ gives the same stability implications as $(F+G)<0$. Hence it is only interesting to look at $(F+G)=0$. Suppose this is true for a discrete time then it doesn't have any impact on physics like the coordinate roots. As for timeintervals with $(F+G)=0$ obviously $D$ behaves like a free particle and thus $L$ will not be stable. Hence $V$ holds (\ref{dV}) as well as $V(L,L)=\dot V(L,L)=0$ with $L:=(\chi,\psi,\zeta,v_\chi,v_\psi,v_\zeta)$, after continuously completing $\dot V$ at non-differentiable points. Thus every region around $(L,L)$ contains points with $V<0$ making the local region unstable and assuring this region to have a radius $R>0$.

\section*{Implication}
\paragraph{}
As for any local region $L$ to be unstable implies that any dense object or region $D$ will leave $L$ in a finite time. $N$-body systems, with $N>2$, have trajectories on which energy efficient radial migration is possible (for calculations see \citet{del}). It is well known that $L$ migrates on such trajectories. Hence $D$ also migrates on such trajectories, but soon leaving $L$ and entering a different local region, which will also travel along such trajectories. Thus $D$ has a high potential of migrating large distances radially through the accreting system $A$. The average distance grows with $R$. This will result in a high mixture of objects within $A$.
\paragraph{} \textbf{Theorem 1}\\
Let $D$ be a dense region or object in an accreting system $A$ with local region $L$, then $L$ is unstable, and $D$ will be migrating radially through $A$. Phenomenologically objects $D$ in $A$ will be close to objects $D'$ born far away in $L'\not=L$.

\end{document}